\documentclass[aps,prb,twocolumn,groupedaddress,amsmath,letterpaper]{revtex4}

\usepackage{graphicx}
\usepackage{array}

\begin{document}

\title{Absorbing boundary condition for Bloch-Floquet eigenmodes}

\author{Chris Fietz}
\email[Email: ]{fietz.chris@gmail.com}
\affiliation{Iowa State University, Ames, Iowa 50011, USA}

\begin{abstract}
We present an absorbing boundary condition for electromagnetic frequency domain simulations of photonic crystals and metamaterials.  This boundary condition can simultaneously absorb multiple Bloch-Floquet eigenmodes of a periodic crystal, including both propagating and evanescent modes.  The photonic crystal or metamaterial in question can include lossy, active, anisotropic and even bi-anisotropic inclusions.  The absorbing boundary condition is dependent on an orthogonality condition for Bloch-Floquet eigenmodes, a generalized version of which is presented here.  We test this absorbing boundary condition numerically and present the results.
\end{abstract}


\maketitle

\section{Introduction}
The numerical simulation of photonic crystals (PCs) has become a very active field over the past twenty years, and this interest has only grown with the emergence of the related field of metamaterials.  A common problem in electromagnetic simulations of PCs and PC waveguides is the termination of a domain inside the crystal.  This requires a boundary condition that can absorb the Bloch eigenmodes of the particular PC or PC waveguide.  One attempt to reduce the reflections from the boundaries of PC waveguides involves terminating the domain with a Distributed Bragg Reflector Waveguide~\cite{Mekis_99} (basically a Bragg Reflector PC waveguide with a wavenumber that matches the original PC waveguide).  There have been several other attempts to solve this problem using Perfectly Matched Layers~\cite{Berenger_94,Chew_94,Jin_02_PML} (PMLs) of different varieties to absorb crystal eigenmodes.  These methods must be distinguished between those that use a simple PML for absorption~\cite{Chen_96,Tsuji_02} and those that apply a PML to the same PC structure that they are attempting to terminate~\cite{Koshiba_01,Kosmidou_03,Weily_03,Oskooi_08,Askari_11}, known as a PCPML.  Although sophisticated applications of a PCPML can terminate boundaries with reflections of $-30\mathrm{dB}$, adding the PCPML to the simulation domain increases the size of the total domain, adding to the computational expense of the simulation.

For electromagnetic eigenmodes that evolve harmonically in the direction of propagation (plane waves, simple waveguides) there is already a well known method for terminating the domain.  It involves imposing a mixed boundary condition relating the derivative of the field in the direction normal to the boundary to the amplitude of the field, $\partial\Psi/\partial n = -\mathrm{i}k\Psi$, where $\Psi$ is the field and $k$ is the wavenumber of the plane wave or waveguide mode.  This boundary condition can be generalized to absorb multiple eigenmodes of the system~\cite{Jin_02_ABC}.  Unfortunately, the requirement that the eigenmode evolve harmonically in the direction normal to the boundary makes this boundary condition inappropriate for a PC, PC waveguide or a metamaterial.  In this paper we generalize the boundary condition in Ref.~\cite{Jin_02_ABC} to absorb multiple Bloch eigenmodes.  In Sec.~\ref{Sec_2} we derive this Bloch mode Absorbing Boundary Condition (Bloch-ABC) for the frequency domain and show how to impose it in a finite element simulation.  In Sec.~\ref{Sec_3} we use the Bloch-ABC to calculate the scattering from an interface between vacuum and a 2D PC.  In Sec.~\ref{Sec_4} we use the Bloch-ABC to terminate a 2D PC waveguide.  Finally, the Bloch-ABC presented in this paper requires an orthogonality relation for Bloch eigenmodes that we derive in Appendix~\ref{solution}.

\section{Bloch Absorbing boundary condition}\label{Sec_2}

When solving for the electric field $\textbf{E}$ of an electromagnetic wave, the wave equation is 

\begin{equation}
\epsilon\displaystyle\frac{\omega^2}{c^2}\textbf{E}-\nabla\times\left(\frac{1}{\mu}\nabla\times\textbf{E}\right) = 0.
\end{equation}

\noindent Here we have assumed that $\epsilon$ and $\mu$ are isotropic, though it is simple to generalize this argument to more complicated materials.  We intend solve the wave equation with a finite element simulation.  For a review of the finite element method used in electromagnetic simulations see Ref.~\cite{Jin_02}.  The weak expression for this wave equation is 

\begin{equation}\label{weak_E}
\mathrm{F_E}(\textbf{v},\textbf{E}) = \epsilon\displaystyle\frac{\omega^2}{c^2}\textbf{v}\cdot\textbf{E}-(\nabla\times\textbf{v})\frac{1}{\mu}(\nabla\times\textbf{E}).
\end{equation}

\noindent Here $\textbf{v}$ is a test function.  Integrating the weak expression over the simulation domain $\Omega$ gives us

\begin{equation}\label{weak_integration}
\begin{array}{rl}
\displaystyle\int_{\Omega} d^3x & \ \ \mathrm{F_E}(\textbf{v},\textbf{E}) = \\[15pt] 
& \displaystyle\int_{\Omega} d^3x \ \ \textbf{v}\cdot \left[ \epsilon\displaystyle\frac{\omega^2}{c^2}\textbf{E} -\nabla\times\left(\frac{1}{\mu}\nabla\times\textbf{E}\right) \right] \\[15pt]
& + \displaystyle\oint_{\partial\Omega} da\ \  \textbf{v}\cdot\left[\hat{\textbf{n}}\times\left(\displaystyle\frac{1}{\mu}\nabla\times\textbf{E}\right)\right],
\end{array}
\end{equation}

\noindent an integral of the wave equation over the simulation domain $\Omega$, and a boundary integral over the domain boundary $\partial\Omega$.  Using the Maxwell equations, the integrand of the boundary integral can be interpreted as $\hat{\textbf{n}}\times(1/\mu\nabla\times\textbf{E}) = -\mathrm{i}\omega/c\hat{\textbf{n}}\times\textbf{H}$.  Setting the integral of the weak expression to zero enforces the wave equation and also imposes a boundary condition through the boundary integral.  Specifically it forces the components of $\textbf{H}$ tangential to the boundary to equal zero.  This \textit{Neumann} boundary condition is the natural boundary condition of the weak expression in Eq.~(\ref{weak_E}) and is known as a \textit{perfect magnetic conductor} boundary condition.  Our objective is to modify this boundary condition on a segment of the simulation boundary so that it becomes a \textit{mixed} boundary condition that absorbs Bloch eigenmodes of the PC.

We do this by borrowing from the previously established boundary condition for absorbing waveguide eigenmodes~\cite{Jin_02_ABC}.  This requires an orthogonality relation for the Bloch eigenmodes of a PC.  The orthogonality relation that we use is a generalization (see Appendix~\ref{solution} for the derivation) of that presented in Refs.~\cite{Johnson_01,Johnson_02,Skorobogatiy_02,Michaelis_03,Song_10}, generalized to work with lossy PCs

\begin{equation}\label{orth}
\int_{\mathrm{B}} d\textbf{n}\cdot \left[\tilde{\textbf{E}}_i^* \times \textbf{H}_j + \textbf{E}_j \times \tilde{\textbf{H}}_i^* \right] = S_i\delta_{ij}.
\end{equation}

\noindent Here $S_i$ is a normalization factor which in general is a complex number, and $\int_{\mathrm{B}} d\textbf{n}\cdot$ is an surface integral over a domain boundary with $d\textbf{n}$ an infinitesimal area vector normal to the boundary.  $\textbf{E}_i$ and $\textbf{H}_i$ are the electric and magnetic fields of eigenmodes of a PC where the eigenvalue is the complex valued wavenumber~\cite{Marcelo_07,Fietz_11}.  $\tilde{\textbf{E}}_i$ and $\tilde{\textbf{H}}_i$ are complementary eigenmodes obtained by solving the complementary eigenvalue problem, this being the original eigenvalue problem with loss replaced with the equivalent amount of gain.  The complex valued wavenumber eigenvalues of the complementary eigenmodes are the complex conjugates of the eigenvalues of the normal eigenmodes (see Appendix~\ref{solution}).  The use of the complementary eigenmodes allows the orthogonality relation to work not only with both lossy and active crystals, but to also with evanescent eigenmodes.  It should be noted that the normalization of the eigenmodes $\textbf{E}_i$ and $\textbf{H}_i$ and their complementary eigenmodes $\tilde{\textbf{E}}_i$ and $\tilde{\textbf{H}}_i$ are arbitrary.  One can multiply $\textbf{E}_i$ and $\textbf{H}_i$ by an arbitrary factor and divide $\tilde{\textbf{E}}_i$ and $\tilde{\textbf{H}}_i$ by the same arbitrary factor and $S_i$ remains unchanged.

We modify the natural boundary condition of the weak expression by specifying the values of the magnetic field tangent to the boundary.  We want the magnetic field at the boundary to be equal to the sum of magnetic fields of Bloch eigenmodes excited at the boundary (modes entering the simulation domain) as well as eigenmodes incident upon the boundary that we require to be absorbed by the boundary (modes exiting the simulation domain).  We set the integrand in the surface integral of Eq.~(\ref{weak_integration}) to

\begin{equation}\label{bound1}
\hat{\textbf{n}}\times\left(\displaystyle\frac{1}{\mu}\nabla\times\textbf{E}\right) = -\mathrm{i}\frac{\omega}{c} \hat{\textbf{n}}\times \left[ \sum_{n+} \mathrm{A}_n^{inc} \textbf{H}_n + \sum_{n-} \mathrm{A}_n^{refl} \textbf{H}_n \right],
\end{equation}

\noindent where $\sum_{n+}$ is a sum over eigenmodes that carry energy into the domain (excited at the boundary) and $\sum_{n-}$ is a sum over eigenmodes that carry energy out of the domain (reflected from within the domain).  The amplitudes for the incident eigenmodes $\mathrm{A}_n^{inc}$ are controlled parameters of the simulation while the amplitudes for the reflected eigenmodes $\mathrm{A}_n^{refl}$ are unknown.  Using the orthogonality relation in Eq.~(\ref{orth}), we represent the unknown reflected eigenmode amplitudes as a functional of the electromagnetic field,

\begin{equation}\label{bound2}
\mathrm{A}_n^{refl} = \displaystyle\frac{\int_\mathrm{B}da\  \left[ \tilde{\textbf{E}}_n^*\times\textbf{H}+\textbf{E}\times\tilde{\textbf{H}}_n^*\right]}{\int_\mathrm{B}da\  \left[ \tilde{\textbf{E}}_n^*\times\textbf{H}_n+\textbf{E}_n\times\tilde{\textbf{H}}_n^*\right]}.
\end{equation}

\noindent Here the index $n$ only includes eigenmodes carrying energy out of the domain.  Combining Eqs.~(\ref{bound1}) and (\ref{bound2}), we derive the weak contribution to the boundary

\begin{equation}
\begin{array}{rl}
\mathrm{B}(\textbf{v},\textbf{E}) = & \textbf{v}\cdot \mathrm{i}\displaystyle\frac{\omega}{c} \left[ \sum_{n+} \mathrm{A}_{n}^{inc} \textbf{H}_n \right.\\[15pt]
& + \displaystyle\left. \sum_{n-}\textbf{H}_n \frac{\int_\mathrm{B}da\  \left[ \tilde{\textbf{E}}_n^*\times\textbf{H}+\textbf{E}\times\tilde{\textbf{H}}_n^*\right]}{\int_\mathrm{B}da\  \left[ \tilde{\textbf{E}}_n^*\times\textbf{H}_n+\textbf{E}_n\times\tilde{\textbf{H}}_n^*\right]} \right].
\end{array}
\end{equation}

\noindent This weak expression must be added to the particular boundary intended to absorb the Bloch eigenmodes of the PC.  We note that although this derivation was for a finite element simulation solving for the electric field $\textbf{E}$, it is simple to adapt it to a simulation solving for the magnetic field $\textbf{H}$.  Throughout this paper all numerical simulations, both those demonstrating the Bloch-ABC as well as any eigenvalue simulations, were performed using Comsol Multiphysics 3.5a.

We end this section by noting that the Bloch-ABC is numerically exact, with unwanted reflections due only to the finite level of discretization, and to a much lesser extent the finite numerical precision of computers.

\section{Example: Two D Photonic Crystal}\label{Sec_3}

\begin{figure}[t]
\begin{center}
\includegraphics[width=\columnwidth]{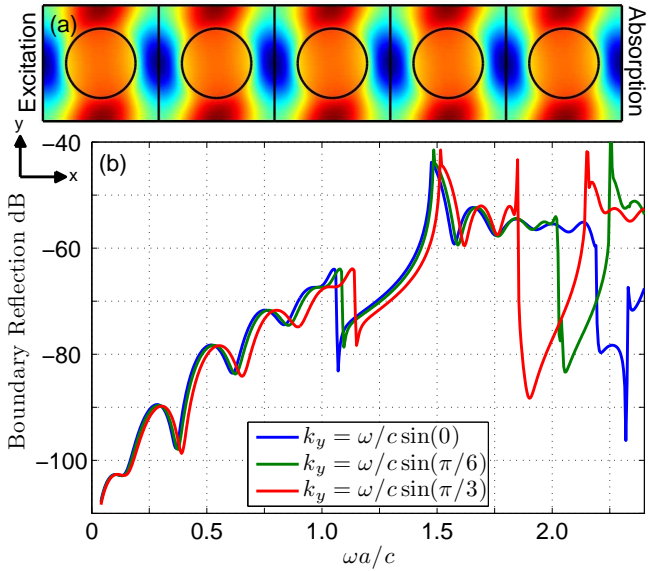}
\end{center}
\caption{(a) $\vert \mathrm{H}_z \vert$ of a $\omega a/c=0.5$, $k_y=\omega/c\sin(\pi/6)$ eigenmode propagating through five unit cells of the PC.  The cylinder in the center of the unit cell is vacuum with permittivity $\epsilon=1$.  The surrounding area is a dielectric with permittivity $\epsilon=5-\mathrm{i}\cdot10^{-6}$.  The eigenmode is excited at the left boundary and absorbed at the right boundary.  The negligible reflection can be seen visually from the absence of interference in the field profile.  (b)  Power reflection from the right boundary normalized to the incident power for three different values of $k_y$.  Both incident and reflected powers were calculated using Eq.~(\ref{Refl}).}\label{Fig_1}
\end{figure}

As an initial example, we consider a simple 2D PC shown in Fig.\ref{Fig_1}(a).  The cylinder in the center of the crystal unit cell has a radius of $0.3\cdot a$ where $a$ is the lattice constant of the unit cell.  This cylinder consists of vacuum with permittivity $\epsilon=1$, and the surrounding area is a dielectric with permittivity $\epsilon=5-\mathrm{i}\cdot 10^{-6}$.  Using the wave convention $e^{\mathrm{i}(\omega t-\textbf{k}\cdot\textbf{x})}$, the negative imaginary part of $\epsilon$ implies loss.

Our first test of the Bloch-ABC is to excite a p-polarized ($\textbf{H}=\mathrm{H}\hat{\textbf{e}}_z$) eigenmode of the PC at one boundary, allow the wave to propagate in the $\hat{\textbf{x}}$ direction as a Bloch mode through a length of PC 5 unit cells long, and then absorb the eigenmode at the terminating boundary.  The crystal is periodic in the $\hat{\textbf{y}}$ direction.  We can measure the reflection from the terminating boundary using a version of the orthogonality relation in Eq.~(\ref{orth})

\begin{equation}\label{Refl}
\mathrm{A}_n = \displaystyle\frac{\frac{1}{2}\int_\mathrm{B}da\  \left[ \tilde{\textbf{E}}_n^*\times\textbf{H}+\textbf{E}\times\tilde{\textbf{H}}_n^*\right]}{\left(\frac{1}{2}\int_\mathrm{B}da\  \left[ \tilde{\textbf{E}}_n^*\times\textbf{H}_n+\textbf{E}_n\times\tilde{\textbf{H}}_n^*\right]\right)^{1/2}}.
\end{equation}

\noindent The amplitude $\mathrm{A}_n$ has units of Power$^{1/2}$, and for propagating modes in lossless crystals, $\vert A_n \vert^2$ can be interpreted as the power flux of the eigenmode through the boundary integrated over in Eq.~(\ref{Refl}).

\begin{figure}[!t]
\begin{center}
\includegraphics[width=\columnwidth]{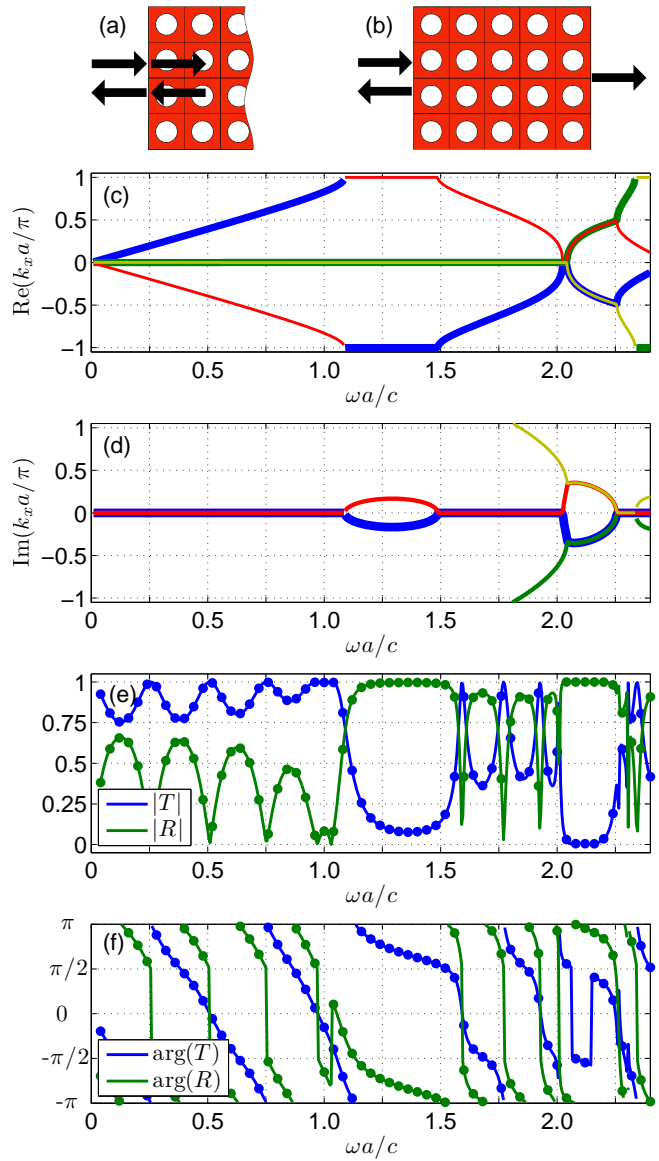}
\end{center}
\caption{(a) Diagram of the scattering simulation.  Plane waves are incident upon the vacuum-PC interface from the vacuum, and Bloch eigenmodes are incident upon the interface from the PC.  The amplitudes of the outgoing waves are measured using Eq.~(\ref{Refl}) and normalized to the incident ampitudes to define the scattering amplitudes of the interface. (b) Diagram of the simulation of reflection and transmission of plane waves through a five layered PC slab. (c) Real and (d) imaginary parts of $k_x(\omega)$ calculated from a complex wavenumber eigenvalue simulation~\cite{Marcelo_07}.  (e) Absolute value and (f) argument (phase) of the transmision and reflection amplitudes returned by the five layer PC slab simulation diagramed in Fig.~\ref{Fig_2}(b) (solid lines) and calculated using the interface scattering amplitudes (dotted lines).}\label{Fig_2}
\end{figure}

The results of this test are shown in Fig.~\ref{Fig_1}(b).  We quantify the efficiency of the Bloch-ABC by calculating the reflection at the Bloch-ABC boundary.  In Fig.~\ref{Fig_1}(b) we plot the power reflected from the Bloch-ABC boundary normalized to the incident power, vs. the normalized frequency $\omega a/c$ with $k_y=\omega/c \sin(\theta)$ for three different angles, $\theta=0,\pi/6,\pi/3$.  We see that in all cases, the reflection is less than $-40\mathrm{dB}$, and we note that this reflection can be further reduced by using a finer mesh.

One application of the Bloch-ABC as well as the orthogonality relation is to calculate the scattering from an interface between a homogeneous medium (vacuum for example) and a periodic crystal.  As a second test of the Bloch-ABC, we have calculated the scattering amplitudes at the interface between vacuum and the PC shown in Fig.~\ref{Fig_1}(a), and then used these amplitudes to calculate the total transmission and reflection from a PC slab five unit cells thick, surrounded on both sides by vacuum.  As can be seen from Fig.~\ref{Fig_2}, the scattering amplitudes of the vacuum-PC interface accurately reproduce the reflection and transmission amplitudes of a five layered PC slab.

We note that in the simulation calculating the scattering amplitudes at the interface between vacuum and the PC, the Bloch-ABC was configured to absorb three Bloch eigenmodes simultaneously.  For most frequencies one eigenmode was propagating and the remaining two were evanescent, though the dispersion curves in Fig~\ref{Fig_2} show multiple propagating modes for frequencies near $\omega a/c=2.5$.  Because the Bloch-ABC was absorbing the most prominent evanescent Bloch eigenmodes (the evanescent eigenmodes with the smallest $\vert \mathrm{Im}(k_x) \vert$), only a single PC unit cell was required in the simulation domain.

\section{Example: Photonic Crystal Waveguide}\label{Sec_4}

\begin{figure}[t]
\begin{center}
\includegraphics[width=\columnwidth]{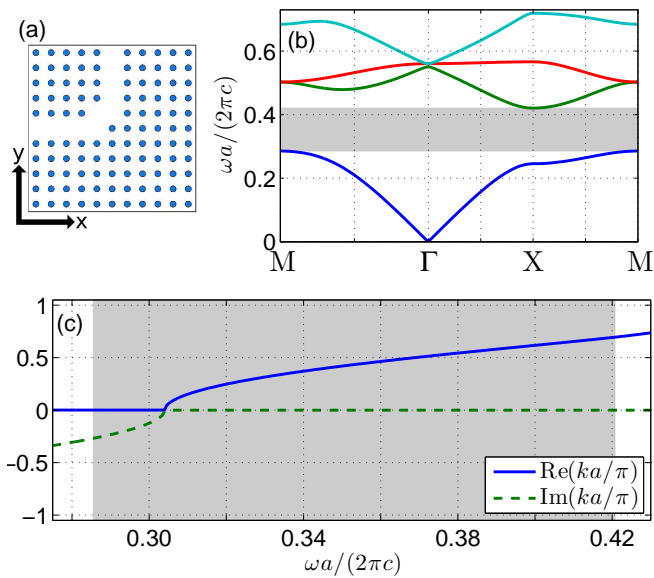}
\end{center}
\caption{(a) Diagram of the 2D PC waveguide simulation, calculating the reflection from a $90^{\circ}$ turn in the PC waveguide.  (b) Band diagram of a $\textbf{E}=\mathrm{E}_z\hat{\textbf{e}}$ eigenmode of the PC.  The bandgap is indicated by the shaded area.  (c) Complex wavenumber dispersion curve~\cite{Marcelo_07} of the PC wavguide eigenmode.  The bandgap of the PC is again indicated by the shaded area.  Note that inside the bandgap there is a cutoff frequency for the PC waveguide mode.  Also, at the high frequency end of the bandgap the PC waveguide mode remains confined to the waveguide even as the frequency exits the bandgap.  At frequencies above the bandgap, there is an additional non-evanescent leaky mode in the PC waveguide that is not shown here.}\label{Fig_3}
\end{figure}

Our second demonstration of the Bloch-ABC involves a PC waveguide.  Fig~\ref{Fig_3}(a) shows a digram of the simulation domain of the PC waveguide.  The PC consists of cylinders with permittivity $\epsilon=11.56$ and radius $r=0.2\cdot a$ where $a$ is the lattice constant of the square unit cell.  The dielectric cylinders are surrounded by vacuum with $\epsilon=1$.  As can be seen from the band diagram plotted in Fig~\ref{Fig_3}(b), this PC has a bandgap for the frequencies $0.286 < \omega a/(2\pi c) < 0.421$.  A PC waveguide is formed by removing a series of individual cylinders along a line.  This waveguide supports an $\textbf{E}=\mathrm{E}_z\hat{\textbf{e}}$ waveguide mode, the dispersion relation of which is plotted in Fig.~\ref{Fig_3}(c).  We will now use the Bloch-ABC to calculate the the reflection of a $90^{\circ}$ turn in the PC waveguide, a problem previously used to demonstrate PCPML schemes~\cite{Koshiba_01,Kosmidou_03}.

\begin{figure}[h]
\begin{center}
\includegraphics[width=\columnwidth]{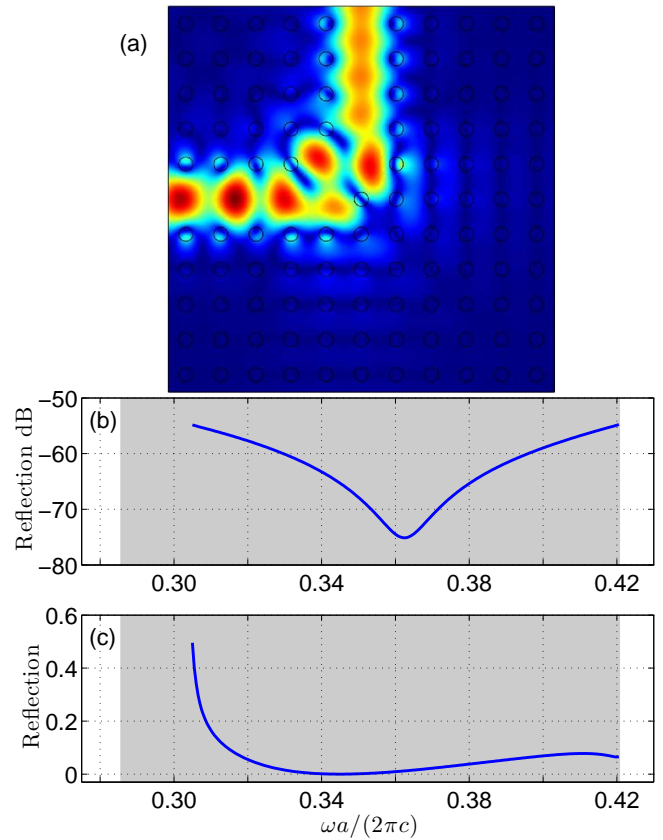}
\end{center}
\caption{(a) $\vert \mathrm{E}_z\vert$ for the PC waveguide bend simulation at frequency $\omega a/(2\pi c)=0.41$.  The waveguide eigenmode is excited at the left boundary of the domain and absorbed at both the left and top boundaries.  Note the interference in the horizontal branch of the waveguide due to the reflection at the $90^{\circ}$ bend and the lack of interference in the vertical branch of the PC waveguide due to negligible reflection at the Bloch-ABC boundary.  (b)  Power reflection from the top Bloch-ABC boundary normalized to power incident upon that boundary.  (c) Power reflection from the $90^{\circ}$ bend in the PC waveguide normalized to power incident upon the $90^\circ$ bend.  The shaded area indicates the PC bandgap shown in Fig~\ref{Fig_3}(b).}\label{Fig_4}
\end{figure}

Using the field profile returned from a complex wavenumber eigenvalue simulation~\cite{Marcelo_07} for the single propagating mode inside the PC waveguide, we can impose a Bloch-ABC on the left and top boundaries of the simulation domain shown in Fig.~\ref{Fig_3}(a).  In addition to absorbing the PC waveguide mode we can use the Bloch-ABC to excite the same mode at the left boundary of the domain.  We then use Eq.~(\ref{Refl}) to calculate the power reflected from the top boundary, quantifying the efficiency of the Bloch-ABC.  In a similar way we calculate the power reflection from the $90^{\circ}$ bend in the PC waveguide, ostensibly the objective of this exercise.  These reflections are plotted in Fig.~\ref{Fig_4}.

From Fig~\ref{Fig_4} we see excellent absorption at the Bloch-ABC.  The reflection from the top Bloch-ABC is always less than $-50\mathrm{dB}$, and could be further reduced with a finer mesh.  In the field profile in Fig~\ref{Fig_4}(a) we see no interference in the vertical branch of the PC waveguide due to essentially no reflection from the top Bloch-ABC.  This is in contrast to the clear interference in the horizontal branch of the PC waveguide due to reflection from the $90^{\circ}$ bend.  Finally, we note that in the simulation shown in Fig~\ref{Fig_4}(a), the Bloch-ABC was configured to absorb only the single propagating waveguide eigenmode.

We end this section by contrasting the Bloch-ABC method of terminating the simulation domain with the use of PCPMLs.  First, the reflection from the Bloch-ABC boundary ($<-50\mathrm{dB}$) is significantly smaller than that from a PCPML~\cite{Koshiba_01,Kosmidou_03}.  Second, the employment of a PCPML requires an additional domain for the PCPML, increasing the overall size of the simulation.  Third, the PCPML is an approximate method for absorbing Bloch eigenmodes while the Bloch-ABC is numerically exact.  The one advantage of a PCPML is that it can be used in a time domain simulation, whereas the Bloch-ABC presented here is limited to the frequency domain.

\section{Conclusion}

We have derived an absorbing boundary condition for the Bloch eigenmodes of periodic structures for use in frequency domain simulations.  We have tested the Bloch-ABC on a PC as well as a PC waveguide, showing that reflection from the Bloch-ABC is negligibly small.  In fact, the Bloch-ABC is numerically exact.  Unwanted reflections can in principle be reduced by refining the mesh.  The Bloch-ABC relies on a orthogonality relation for Bloch eigenmodes~\cite{Johnson_01,Johnson_02,Skorobogatiy_02,Michaelis_03,Song_10} that we have generalized (in Appendix\ref{solution}) for a wider variety of periodic media, including lossy media.

\section*{Acknowledgements}
This work was supported by the IC Postdoctoral Research Fellowship Program.  I would also like to acknowledge the influence of Peng Zhang, whose own work required the Bloch-ABC, and therefore served as the impetus for its development.

\appendix
\section{Orthogonality relation for Bloch eigenmodes}\label{solution}

A necessary requirement for the Bloch-ABC is an orthogonality relation between different Bloch eigenmodes of a PC, PC waveguide or metamaterial.  An orthognality relation for Bloch eigenmodes has previously been presented in several papers~\cite{Johnson_01,Johnson_02,Skorobogatiy_02,Michaelis_03,Song_10}.  Here we generalize this orthogonality relation to work with eigenmodes of lossy (and active) crystals, as well as crystals made up of anisotropic and bianisotropic materials.  This generalized orthogonality relation also works with evanescent eigenmodes.  Aside from this generalization, our derivation of the orthogonality relation closely follows that of Ref.~\cite{Song_10}.

The electric and magnetic fields of a crystal Bloch eigenmode are represented as

\begin{equation}
\begin{array}{c}
\textbf{E}(t,\textbf{x}) = \textbf{e}(\textbf{x}) e^{\mathrm{i}(\omega t-\textbf{k}\cdot\textbf{x})}, \\
\textbf{H}(t,\textbf{x}) = \textbf{h}(\textbf{x}) e^{\mathrm{i}(\omega t-\textbf{k}\cdot\textbf{x})},
\end{array}
\end{equation}

\noindent where $\textbf{e}$ and $\textbf{h}$ are periodic vectors 

\begin{equation}
\begin{array}{c}
\textbf{e}(\textbf{x}+\textbf{N}\cdot\textbf{a}) = \textbf{e}(\textbf{x}), \\
\textbf{h}(\textbf{x}+\textbf{N}\cdot\textbf{a}) = \textbf{h}(\textbf{x}),
\end{array}
\end{equation}

\noindent where $\textbf{a}$ is the set of lattice vectors for the unit cell of the PC, and $\textbf{N}$ is a vector of integers.  The Bloch wavevector $\textbf{k} = \textbf{k}_0 + \beta\hat{\textbf{n}}$ can be divided into a constant part $\textbf{k}_0$ and a variable part $\beta\hat{\textbf{n}}$, both parts potentially being complex valued.  Here $\hat{\textbf{n}}$ is a unit vector and $\beta$ is the eigenvalue we will use to prove the orthogonality relation.

Using these periodic vector functions we can represent the Maxwell equations as two separate eigenvalue problems

\begin{equation}
\begin{array}{c}
\left(\hat{A}-i\hat{B}\displaystyle\frac{\partial}{\partial n}\right)\cdot\psi_i = \beta_i\hat{B}\psi_i, \\ \\
\left(\hat{A}-i\hat{B}\displaystyle\frac{\partial}{\partial n}\right)^\dagger\cdot\tilde{\psi}_i = \beta_i^*\hat{B}\tilde{\psi}_i,
\\ \\
\hat{A} = 
\left(\!\!\!\begin{array}{cc}
\hat{\epsilon}\displaystyle\frac{\omega}{c} & \hat{\xi}\displaystyle\frac{\omega}{c}+\mathrm{i}\nabla_t\times+\textbf{k}_0\times \\
\hat{\zeta}\displaystyle\frac{\omega}{c}-\mathrm{i}\nabla_t\times-\textbf{k}_0\times & \hat{\mu}\displaystyle\frac{\omega}{c}
\end{array}\!\!\!\right),
\\ \\
\hat{B} = 
\left(\!\!\!\begin{array}{cc}
0 & -\hat{\textbf{n}}\times \\
\hat{\textbf{n}}\times & 0
\end{array}\!\!\!\right),
\\ \\
\psi_i = 
\left(\!\!\!\begin{array}{c}
\textbf{e}_i \\ \textbf{h}_i
\end{array}\!\!\!\right),
\\ \\
\tilde{\psi}_i = 
\left(\!\!\!\begin{array}{c}
\tilde{\textbf{e}}_i \\ \tilde{\textbf{h}}_i
\end{array}\!\!\!\right).
\end{array}
\end{equation}

\noindent Here $\nabla_t\equiv\nabla-\hat{\textbf{n}}(\hat{\textbf{n}}\cdot\nabla)$ is the $\nabla$ operator projected onto a plane tangent to the unit vector $\hat{\textbf{n}}$ while $\partial/\partial n \equiv \hat{\textbf{n}}\cdot\nabla$ is the $\nabla$ operator projected onto the unit vector $\hat{\textbf{n}}$.  $\psi_i$ are the Bloch eigenmodes and $\tilde{\psi}_i$ are the complementary Bloch eigenmodes, both with the Bloch phase factor divided off.  The complementary eigenmodes are the modes that would result from replacing the loss in a PC with an equivalent amount of gain.  $\tilde{\psi}_i^{\dagger}$ and $\psi_i$ are sometimes called the left and right generalized eigenfunctions of the pair of operators $\hat{A}$ and $\hat{B}$.  Appendix~\ref{lr_eigen} contains a short primer on the orthogonality of left and right generalized eigenvectors.  Note that the sets of eigenvalues for the normal and complementary eigenvalue problems are complex conjugates of each other.  Also, $\hat{B}$ is always hermitian.  In the case of a lossless PC, if $\textbf{k}_0$ is restricted to be real valued, $\hat{A}$ is hermitian, making the operator $\hat{A}-i\hat{B}\partial/\partial n$ hermitian.  In this lossless case, the eigenvalues $\beta_i$ are either real valued (propagating modes) or they come in complex conjugate pairs (evanescent modes).

We define the following inner product

\begin{equation}\label{inner_prod2}
\langle \tilde{\psi}_i |\hat{O}| \psi_j \rangle \equiv \int_{\Omega} dn 
\left(\!\!\!\begin{array}{c}
\textbf{e}_i \\[3pt] \textbf{h}_i
\end{array}\!\!\!\right)^\dagger\cdot \hat{O}\cdot
\left(\!\!\!\begin{array}{c}
\textbf{e}_j \\[3pt] \textbf{h}_j
\end{array}\!\!\!\right),
\end{equation}

\noindent where $\hat{O}$ is an arbitrary operator and the integral is over a plane of the unit cell perpendicular to the unit vector $\hat{\textbf{n}}$.

Borrowing from the orthogonality proof in Ref.~\cite{Song_10}, by partial integration we can show that 

\begin{equation}
\begin{array}{rl}
\langle\tilde{\psi}_i | \left(\hat{A}-\mathrm{i}\hat{B}\displaystyle\frac{\partial}{\partial n}\right) | \psi_j\rangle = &
\langle\psi_j | \left(\hat{A}-\mathrm{i}\hat{B}\displaystyle\frac{\partial}{\partial n}\right)^\dagger | \tilde{\psi}_i\rangle^* \\[15pt]
& -\mathrm{i}\displaystyle\frac{\partial}{\partial n} \langle \tilde{\psi}_i | \hat{B} | \psi_j\rangle,
\end{array}
\end{equation}

\noindent implying

\begin{equation}
\displaystyle\frac{\partial}{\partial n} \langle\tilde{\psi}_i | \hat{B} | \psi_j\rangle = -\mathrm{i}\left(\beta_i-\beta_j\right) \langle\tilde{\psi}_i | \hat{B} | \psi_j\rangle.
\end{equation}

\noindent The solution to this simple differential equation is $\langle\tilde{\psi}_i | \hat{B} | \psi_j\rangle_n = e^{-\mathrm{i}(\beta_i-\beta_j)n} \langle\tilde{\psi}_i | \hat{B} | \psi_j\rangle_{n=0}$, where $n$ is a displacement in the direction of $\hat{\textbf{n}}$.  However, since $\tilde{\psi}_i$ and $\psi_j$ are Bloch eigenmodes divided by the Bloch phase factor, $\langle\tilde{\psi}_i | \hat{B} | \psi_j\rangle_{n} = \langle\tilde{\psi}_i | \hat{B} | \psi_j\rangle_{n=0}$, implying

\begin{equation}\label{inner_prod_1}
\langle\tilde{\psi}_i | \hat{B} | \psi_j\rangle = \int_{\Omega} d\textbf{n}\cdot \left[\tilde{\textbf{e}}_i^* \times \textbf{h}_j + \textbf{e}_j \times \tilde{\textbf{h}}_i^* \right] = s_i\delta_{ij},
\end{equation}

\noindent where $s_i$ is a normalization factor which in general is a complex number.  Since the integral in Eq.~(\ref{inner_prod_1}) is over dimensions perpendicular to $\hat{n}$ we can simplify the orthogonality relation by using the full field values

\begin{equation}
\int_{\Omega} d\textbf{n}\cdot \left[\tilde{\textbf{E}}_i^* \times \textbf{H}_j + \textbf{E}_j \times \tilde{\textbf{H}}_i^* \right] = S_i\delta_{ij},
\end{equation}

\noindent where again $S_i$ is a complex valued normalization factor.

\section{Primer on orthogonality with a non-hermitian generalized eigenvalue problem}\label{lr_eigen}
When considering the eigenfunctions and complementary eigenfunctions used in the derivation of the orthogonality relation derived in Appendix~\ref{solution}, it may help to make an analogy to linear algebra, replacing eigenfunctions with eigenvectors and replacing differential operators with matrices.  Consider the generalized eigenvalue problem

\begin{equation}
\hat{A}\cdot\textbf{u}_i = \lambda_i\hat{B}\cdot\textbf{u}_i.
\end{equation}

\noindent Here $\hat{A}$ and $\hat{B}$ are $N\times N$ matrices and $\textbf{u}_i$ is a $N\times 1$ vector.  $\textbf{u}_i$ is called a right generalized eigenvector of the pair of matrices $\hat{A}$ and $\hat{B}$.  The left generalized eigenvalue problem

\begin{equation}
\textbf{v}_i\cdot\hat{A} = \lambda_i\textbf{v}_i\cdot\hat{B},
\end{equation}

\noindent defines the left generalized eigenvector $\textbf{v}_i$.  It is important to note that both generalized eigenvalue problems have the same set of eigenvalues $\lambda_i$.  It is easy to see that this is true by considering that both the left and right generalized eigenvalue problems have the same characteristic polynomial $\det(\hat{A}-\lambda\hat{B})$, the zeros of which are the set of eigenvalues $\lambda_i$.

Using these definitions of the generalized eigenvectors, we can easily prove orthogonality between left and right eigenvectors of differing eigenvalues with the equality

\begin{equation}\label{vec_orth}
\textbf{v}_i\cdot \left(\hat{B}\cdot\textbf{u}_j\right) = \displaystyle\frac{1}{\lambda_j} \left(\textbf{v}_i\cdot\hat{A}\right) \cdot\textbf{u}_j = \displaystyle\frac{\lambda_i}{\lambda_j}\textbf{v}_i\cdot\hat{B}\cdot\textbf{u}_j.
\end{equation}

\noindent If $\lambda_i \neq \lambda_j$, then Eq.~(\ref{vec_orth}) implies $\textbf{v}_i\cdot\hat{B}\cdot\textbf{u}_j = 0$.  If $\lambda_i=\lambda_j$, then $\textbf{v}_i\cdot\hat{B}\cdot\textbf{u}_j$ can be nonzero.  We are ignoring the possibility of degenerate eigenvalues, which if encountered can be handled in the usual way by manually orthogonalizing the associated left and right eigenvectors.  The final result is

\begin{equation}
\textbf{v}_i\cdot\hat{B}\cdot\textbf{u}_j = b_i\delta_{ij},
\end{equation}

\noindent where in general $b_i$ is a set of complex numbers.  This equation in function/operator form gives us Eq.~(\ref{inner_prod_1}).  A similar argument allows us to show that 

\begin{equation}
\textbf{v}_i\cdot\hat{A}\cdot\textbf{u}_j = a_i\delta_{ij}.
\end{equation}

\noindent where again in general $a_i$ is a set of complex numbers.  Finally, we note that at no point have we assumed that $\hat{A}$ and $\hat{B}$ were hermitian.


\end{document}